# Asymmetric variation of a finite mass harmonic like oscillator

Jihad Asad[a,*], P. Mallick[b,*], M.E. Samei[c], B. Rath[b,*], Prachiparava Mohapatra[b], Hussein Shanak[a], Rabab Jarrar[a]

[a] *Department of Physics, College of Applied Sciences, Palestine Technical University- Kadoorie (PTUK), Tulkarm, Palestine*
[b] *Department of Physics, North Orissa University, Takatpur, Baripada 757003, Odisha, India*
[c] *Department of Mathematics, Faculty of Basic Science, Bu-Ali Sina University, Hamedan, Iran*

ABSTRACT

Classical and quantum mechanical analysis have been carried out on harmonic like oscillator with asymmetric position dependent mass. Phase space analysis are performed both classically and quantum mechanically for a plausible understanding of the subject.

## Introduction

Position dependent mass (PDM) attracted the attention of many researchers due to their applications in semiconductor physics [1-4], quantum dots [5], quantum liquids [6], nonlinear oscillators [7-9]. Moreover, one can find the formalism of PDM in many other branches of Physics such as Molecular Physics [10], Nuclear Physics [11], quantum information [12] etc. Referring to literature, one can also see that the problem associated with PDM has attracted many researchers and still found wide spread applications in various branches of Physics [12-27].

The non-commutivity of momentum and position imposes on researchers to pay great attention in defining the kinetic energy operator on PDM as well as choosing suitable combination of position and momentum at classical level [28,29]. Much of the works on the position dependent symmetric mass systems have been studied with singular mass of the type [30-32]

$$m(x) = \frac{m_0}{1 - \lambda x^2} \qquad (1a)$$

and mass without singularity [29-33] as

$$m(x) = \frac{m_0}{1 + \lambda x^2} \qquad (1b)$$

The PDM of the form

$$m(x) = 1 + \lambda x^N \qquad (1c)$$

with $N$ = even number has also been reported [34]. Marques et al. [19] studied the singular mass PDM using Dirac equation and reported the presence of cosmic string. In a recent work [14], the PDM creation and annihilation operators constructed for harmonic oscillators through two methods: the PDM point canonical transformation, and the von Roos PDM- Hamiltonian. The author [14] suggested a simple PDM without singularities as the form suggested by Eq. (1b) and a simple PDM with singularities of the form

$$m(x) = (\lambda x^2 - 1)^{-3} \qquad (1d)$$

Further, a nonlinear oscillator with PDM has been recently studied [35] in the approach of supersymmetric quantum mechanics. The authors suggested the following familiar PDM function

$$m(x) = \lambda(1 + \delta^2 x^2)^{-1} \qquad (2)$$

and determined the coherent states for a particular variable mass system in the sense of Barut-Girardello and then compare their properties to those already known. A new type of symmetric solitonic mass of the type [20,26]

$$m(x) = m_0 \sec h^2(\alpha x) \qquad (3)$$

has been reported to give exact energy and wavefunction. The authors have also discussed Shannon information entropy relation. Dong and Lozada-Cassou [23] reported the exact solution for PDM with hard core potential.

However, the PDM under asymmetric has been reported by few. For example: The PDM reflecting asymmetry of the form

$$m(x) = \frac{m_0}{(1 + \gamma x)^2} \tag{4}$$

has been reported in ref. [18]. In a very recent work on PDM applicable to semiconductor heterostructures was developed by El-Nabulsi [22] with the form of $m(x)$ as

$$m(x) = m(1 + \gamma x)^k \tag{5}$$

This new approach resulted into modified momentum operator, modified commutation and uncertainty relation that leads to estimation of density of states in terms of the conventional value and the length of the quantum nanowire. Dong et al. [24] studied the asymmetric model singular mass oscillator with mass of the type

$$m(x) = \frac{1}{\tau^\alpha (x + a)^\alpha} \tag{6}$$

using Lie algebra to obtain its eigenvalues and eigenfunctions. The authors in [28] suggested a PDM of the form

$$m(x) = m e^{ax + \frac{1}{2}bx^2} \tag{7}$$

The PDM stated above is of asymmetric in nature. In fact, it has been seen that mass has asymmetric function of position $(x)$ in semiconductor physics. If $a$ and $b$ are negative then the (Eq. (7)) will be 0 at large values of $x$. In other words, a massive $m(x)$ particle undergoes a drastic change and becomes a photon, which can hardly interact with any matter. Secondly, if the mass behaves as:

$$m(x) = m(1 + x^2) \tag{8}$$

Then at large distance mass becomes infinite. If mass becomes infinite while moving with distance then it becomes meaningless to discuss Physics behind an infinite mass.

**Asymmetric variation of finite mass**

From the review of literature, we have seen that the variation of mass with position is not within the finite limit. In order to avoid the physically unaccepted two extreme cases explained above, we choose a PDM of the form:

$$m(x) = \frac{m}{1 + e^{-x - \lambda x^2}} \tag{9}$$

where the parameter $\lambda$ is small positive constant and considered here as 0.1. The purpose of considering this mass expression in addition to the asymmetric variation of mass with pos $m(x = 0) = m_1 = \frac{m}{2}$ $m(x \to \infty) = m_2 = m$ ition, it is confined within two finite limits such as (i) and (ii) (Fig. 1). Further, the $m(x)$ increases exponentially from $\frac{m}{2}$ to saturate at $m$ as we move away from 0 to positive side. However, the $m(x)$ showed the parabolic shape with distance if we move in opposite direction which finally saturates at $\frac{m}{2}$. This asymmetric nature of the mass is new to the literature. According to the above considerations, we focus our attention on a Harmonic like oscillator with finite and simple asymmetry position dependent mass PDM and investigate the system classically as well as quantum mechanically.

The rest of this work is organized as follows: In Sec. III a classical analysis for the problem is discussed both analytically and numerically. In Sec. IV quantum mechanical analysis is presented, and finally the important findings of the work are outlined in section V.

**Classical analysis**

The Hamiltonian of the oscillator considered here is

$$H = \frac{p_x^2}{2m(x)} + \frac{m(x)}{2} x^2 \tag{10a}$$

The Lagrangian is related to the Hamiltonian as:

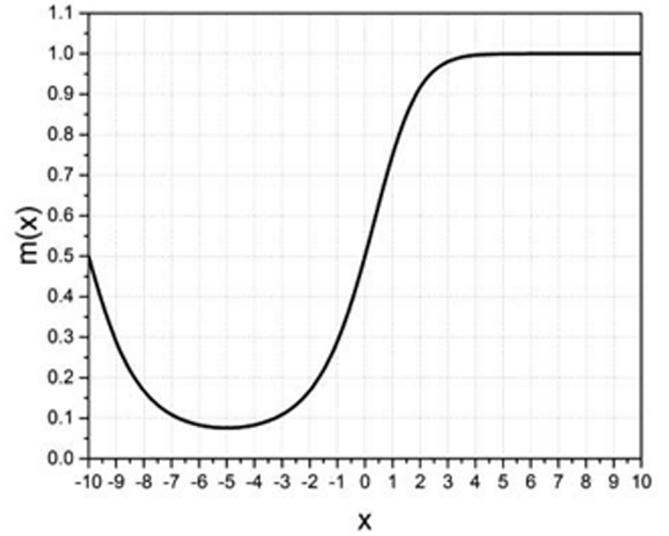

**Fig. 1.** The variation of the mass (Eq. (9)) with position (x) with $\lambda = 0.1$ and $m = 1$.

$$H = \sum_i p_i \dot{q}_i - L \tag{10b}$$

In our case we have just one generalized coordinate$(x)$, and one generalized momenta $(p_x)$. Therefore we have

$$H = p_x \dot{x} - L$$

$$\frac{p_x^2}{2m(x)} + \frac{m(x)}{2} x^2 = p_x \dot{x} - L \tag{10c}$$

Rearrangement of Eq. (10c), gives an expression of Lagrangian ($L$) as:

$$L = p_x \dot{x} - \frac{p_x^2}{2m(x)} - \frac{m(x)}{2} x^2 \tag{10d}$$

Using $\frac{dx}{dt} = \frac{\partial H}{\partial p_x} \Rightarrow \dot{x} = \frac{p_x}{m(x)}$, and $p_x = \dot{x} m(x)$. Thus Eq. (10d) becomes:

$$L = \frac{m(x)}{2}(\dot{x}^2 - x^2) \tag{11}$$

On the substitution of Eq. (11) to the following relation

$$\frac{\partial L}{\partial x} - \frac{d}{dt}\left(\frac{\partial L}{\partial \dot{x}}\right) = 0, \tag{12a}$$

would lead to the equation of motion (i.e. known as: Euler Lagrange Equation) as

$$-xm(x) + \frac{(\dot{x}^2 - x^2)}{2}\frac{dm(x)}{dx} - \ddot{x}m(x) - \dot{x}\frac{dm(x)}{dt} = 0 \tag{12b}$$

Substituting the expression of mass (Eq. (9)) into Eq. (12b), we get the following equation of motion, which in a simplified form can be written as

$$\ddot{x} + x - \left(\frac{\dot{x}^2 + x^2}{2}\right)\left(\frac{1 + 2\lambda x}{1 + e^{-x - \lambda x^2}}\right)e^{-x - \lambda x^2} = 0 \tag{13}$$

It is worth to mention here that Eq. (13) can be derived without using the PDM approach, by using the approach presented in the work of Musielak and others [32,36-41].

*Analytical study*

In order to solve the above equation (Eq. (13)), we follow the He's frequency formulation on ancient Chinese method [42-44] to get the frequency of oscillation using the following condition:

$$\omega^2 = \frac{\omega_1^2 R_2(0) - \omega_2^2 R_1(0)}{R_2(0) - R_1(0)} \qquad (14a)$$

where

$$R_1(0) = -\left(\frac{A}{2}\right)\frac{(1 + 2\lambda A)}{(1 + e^{-A-\lambda A^2})}e^{-A-\lambda A^2} \qquad (14b)$$

and

$$R_2(0) = (1 - \omega_2^2)A - \left(\frac{A}{2}\right)\frac{(1 + 2\lambda A)}{(1 + e^{-A-\lambda A^2})}e^{-A-\lambda A^2} \qquad (14c)$$

In the above formalism, the boundary condition for the solution $x(t)$ will be $x(t = 0) = 1$ and $\dot{x}(t = 0) = 0$. We therefore considered

$$x(t) = A\cos\omega t \qquad (15)$$

In order to get the expression for frequency of oscillation following the He's formalism [42-44] as stated above, one need to substitute $\omega_1 = 1$ and $\omega_2 = \omega$ in the final expression of Eq. (14a). We therefore obtain the frequency of oscillation for the presem$(x) = m = 1$nt case as

$$\omega = \sqrt{1 - \frac{A}{2}\left(\frac{1 + 2\lambda A}{1 + e^{-A-\lambda A^2}}\right)e^{-A-\lambda A^2}} \qquad (16)$$

Fig. 2(a) showed the variation of $\omega$ with $A$ for $\lambda = 0.1$. It can be seen from the figure that for large $A$, $\omega$ is 1. However, for small value of $A$, the value of $\omega$ is less than 1. For large $A$ means large decrease in exponential term to 0 i.e. $e^{-A-\lambda A^2} \approx 0$ which justifying the harmonic oscillator having mass, . The entire variation of frequency with amplitude reflects the asymmetric nature. Fig. 2(b) reflects the variation of $x(t)$ with time ($t$) at different amplitudes i.e. $A = 1, 5, 10$. For $A = 1$, it is just like a simple oscillator having minimum value of frequency $\sim 0.92$. However, for large values of $A$, frequency approached to 1 like that of a simple harmonic oscillator. In other word, the different nature of graph is resulting from the different values of frequency. Fig. 3 reflects the variation of $p(t)$ with $t$ at different amplitudes as was done for $x(t)$. Here, $p(t) \sim A\omega\sin\omega t$ and the sharp rise in the graph (Fig. 3) is due to the contribution of amplitude. For smaller amplitude, it shows oscillating type behaviour whereas the same showed complicated feature at higher amplitude. The trajectory of phase space ($p$ vs $x$) is shown in Fig. 4 and we find the phase space is closed with the reflection of asymmetry. The asymmetry nature is prominent for larger amplitude.

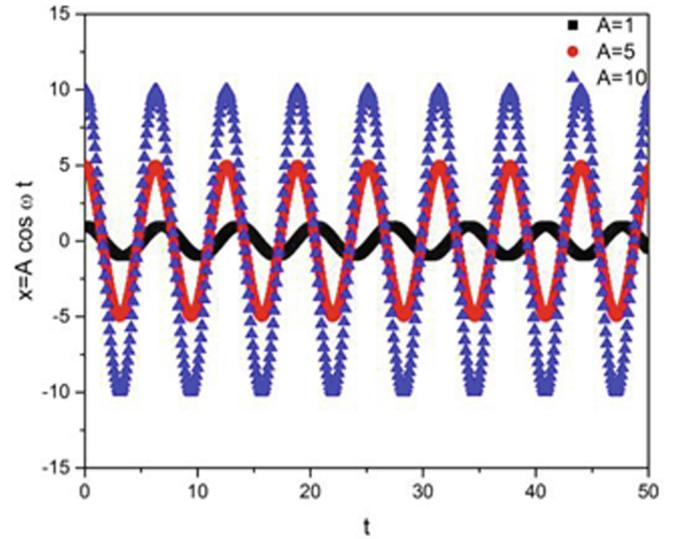

Fig. 2b. Variation of $x(t)$ with time ($t$) for different amplitude.

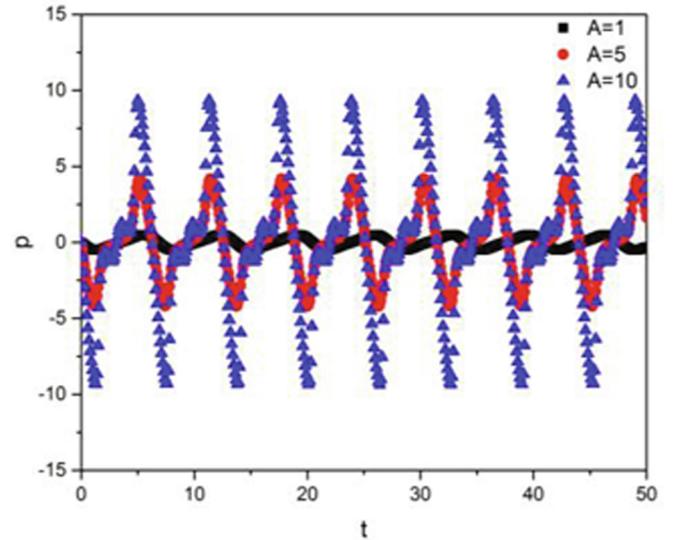

Fig. 3. Variation of $p(t)$ with time ($t$) for different amplitudes.

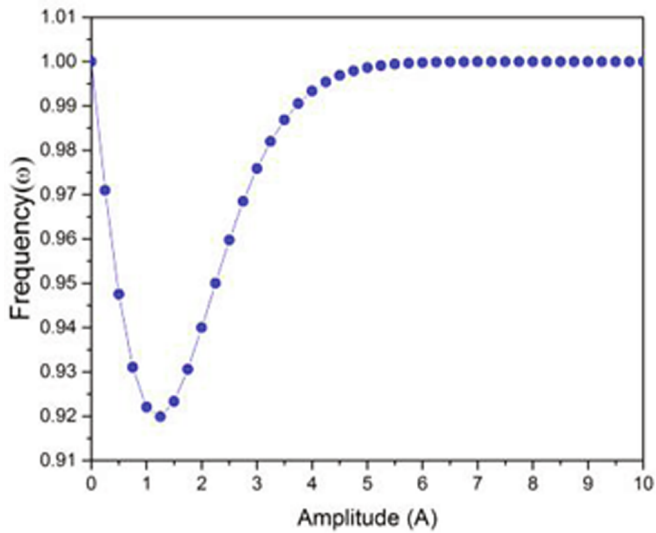

Fig. 2a. Variation of $\omega$ with $A$ for $\lambda = 0.1$.

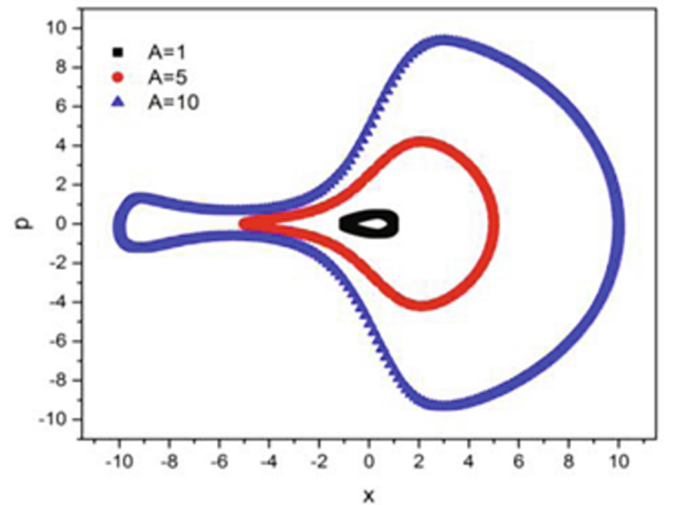

Fig. 4. Trajectory of classical phase space obtained analytically for different amplitude.

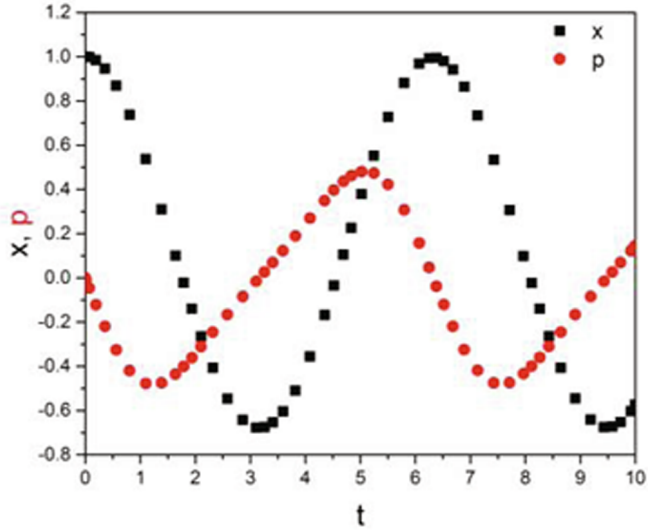

**Fig. 5.** Variation of × and p obtained numerically for $\lambda = 0.1$ with respect to time t.

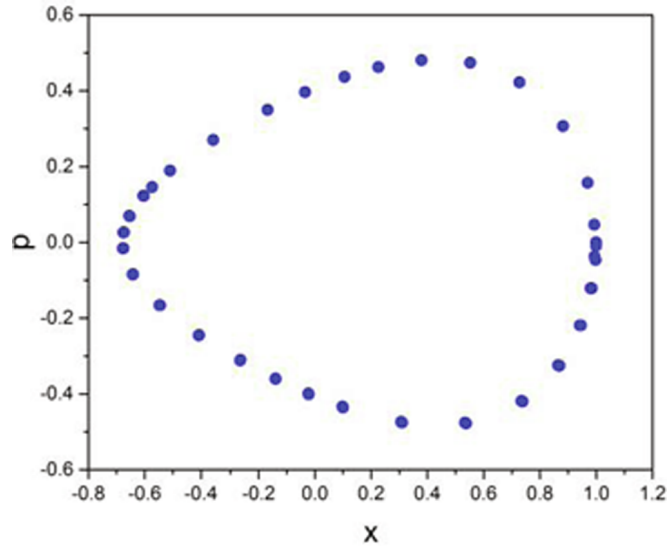

**Fig. 6.** Variation of p obtained numerically for $\lambda = 0.1$ with respect to x.

**Table 1**
Eigenvalues of Asymmetry mass Harmonic like oscillator.

| Quantum number (n) | Computed Eigenvalues |
| --- | --- |
| 0 | 0.5 |
| 1 | 1.2789 |
| 2 | 2.2610 |
| 3 | 3.3735 |
| 4 | 4.5412 |

*Numerical study*

In order to get better information on the classical dynamics of the system (Eq. (13)), the numerical study on the said system is undertaken. We can rewrite Eq. (13) as:

$$\ddot{x} - Q\dot{x}^2 - Qx^2 + x = 0 \tag{17}$$

where

$$Q = \frac{1}{2}\left(\frac{1+2\lambda x}{1+e^{-x-\lambda x^2}}\right)e^{-x-\lambda x^2} = \frac{1+2\lambda x}{2(1+e^{x+\lambda x^2})} \tag{18}$$

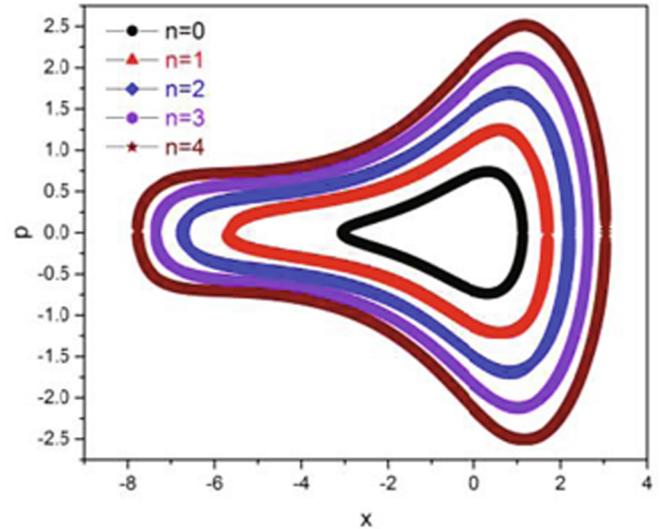

**Fig. 7.** Trajectory of quantum phase space.

Here, we also use the initial condition $x(t=0) = 1$ and $\dot{x}(t=0) = 0$ for the value of $\lambda = 0.1$. Using the algorithm (Appendix-A), we find the numerical solution of Eq. (17) (Table A1). The variation of $x$ and $p$ with respect to time $t$ and phase space trajectory ($p$ with respect to $x$) obtained numerically for $\lambda = 0.1$ are shown in Figs. 5 and 6 respectively. Here, we also notice the closed path phase space trajectory (Fig. 6) reflecting the identical asymmetric nature as was seen for the analytical case (Fig. 4).

**Quantum mechanical analysis**

Now we analyze the above problem quantum mechanically. In quantum analysis, we solve the eigenvalue relation [34,45,46] as

$$H|\Psi_n\rangle = E|\Psi\rangle \tag{19a}$$

where

$$|\Psi\rangle = \sum_n A_n |\varphi_n\rangle \tag{19b}$$

and $|\varphi_n\rangle$ satisfies the eigenvalue relation for position independent harmonic oscillator as

$$H_0|\varphi_n\rangle = [p^2 + x^2]|\varphi_n\rangle = (2n+1)|\varphi_n\rangle \tag{20a}$$

In the above, $|\varphi_n\rangle$ is the wave function of the simple harmonic oscillator with the form

$$|\varphi_n\rangle = Ae^{-\frac{x^2}{2}}H_n(x) \tag{20b}$$

where $H_n(x)$ is the Hermite polynomial and $A$ is the normalization constant. Further, the Hamiltonian (Eq. (10a)) can be rewritten as [16,34]

$$H = p_x \frac{1}{2m(x)} p_x + \frac{m(x)}{2} x^2 \tag{21}$$

Using the above formalism [34,45,46], we get the recurrence relation as

$$\sum_{k=0,1,2,3,\ldots} P_n^k A_{n-k} + S_n A_n + R_n^k A_{n+k} = 0 \tag{22}$$

where

$$P_n^k = \langle n| H |n-k\rangle \tag{23}$$

$$R_n^k = \langle n| H |n+k\rangle \tag{24}$$

$$S_n = \langle n| H |n\rangle - E \tag{25}$$

The eigenvalues of the Harmonic like oscillator Hamiltonian (Eq. (10a)) with position dependent mass (Eq. (9)) are obtained following the above mentioned procedure for $\lambda = 0.1$ and $m = 1$ (Table 1). A quantum mechanical plot of phase space is illustrated in (Fig. 7) for this case considering $E = H$ of respective states and the nature of the curve clearly indicated the asymmetry involved in the mass. Comparing the trajectory of quantum phase space (Fig. 7) with that of classical phase space obtained both analytically (Fig. 4) and numerically (Fig. 6), we noticed a great similarity between classical and quantum nature.

**Conclusion**

We develop a new modeled Harmonic like oscillator Hamiltonian in which mass of the particle varies asymmetrically between two finite values. This PDM system was studied by using classical (both analytically and numerically) and quantum mechanical approaches. The phase space trajectory of the system showed closed loop reflecting asymmetric nature for each case. The asymmetry in the trajectory of phase space could be due to asymmetry associated the PDM. We believe that the proposed asymmetric mass could find its applicability in semiconductor Physics or other branches of Physics where asymmetry plays an important role.

**CRediT authorship contribution statement**

**Jihad Asad:** Visualization, Investigation, Writing - review & editing, Validation. **P. Mallick:** Data curation, Investigation, Writing - original draft, Supervision, Writing - review & editing, Validation. **M.E. Samei:** Data curation, Software, Validation. **B. Rath:** Conceptualization, Methodology, Supervision, Writing - review & editing, Validation. **Prachiparava Mohapatra:** Writing - review & editing, Validation. **Hussein Shanak:** Writing - review & editing, Validation. **Rabab Jarrar:** Writing - review & editing, Validation.

**Declaration of Competing Interest**

The authors declare that they have no known competing financial interests or personal relationships that could have appeared to influence the work reported in this paper.

**Acknowledgments**

The authors (Jihad Asad, Hussein Shanak, and Rabab Jarrar) would like to thank Palestine Technical University- Kadoorie. The author (M. E. Samei) was supported by Bu- Ali Sina University.

**Appendix-A: The proposed method for to find the solution for Eq. (17) using MATLAB**

```
1. function [MatrixEq] = SolveMainEq(lambda, x0, xprime0, m0, tmin, tmax)
2. % gamma : [gamma];
3. % Eq. 1(r, c): results of main equation;
4. [xlambda, ylambda] = size(lambda);
5. column = 2;
6. for j = 1:ylambda
7. f1 = @(t, x)[x(2); (1 + 2_lambda(j)_x(1))_(x(2))^2/(2_(1 + exp(x(1) + …
    lambda(j)_(x(1))^2))) + …
    (1 + 2_lambda(j)_x(1))_(x(1))^2/(2_(1 + exp(x(1) + …
    lambda(j)_(x(1))^2))) - x(1)];
8. [tr, xr] = ode23(f1, [tmin tmax], [x0, xprime0]);
9. maxrow = max(size(tr));
10. for k = 1:maxrow
11. MatrixEq(k, 1) = k;
12. MatrixEq(k, column) = tr(k);
13. MatrixEq(k, column + 1) = round(xr(k, 1),6);
14. MatrixEq(k, column + 2) = round(xr(k, 2),6);
15. MatrixEq(k, column + 3) = round((1 + 2_lambda(j)_xr(k, 1))/(2_(1 + …
    exp(xr(k, 1) + lambda(j)_(xr(k, 1))^2))),6);
16. mx = round(m0/(1 + exp((-1)_xr(k, 1) - lambda(j)_(xr(k, 1))^2)),6);
17. MatrixEq(k, column + 4) = mx;
18. px = round(mx_xr(k, 2),6);
19. MatrixEq(k, column + 5) = px;
20. MatrixEq(k, column + 6) = round(px^2/(2_mx) + mx_xr(k, 1)/2,6);
21. end;
22. column = column + 7;
23. end;
24. end
```

**Table A1**
Numerical results of Eq. (17) for $\lambda = 0.1$ and $t \in [0, 10]$.

| Sl. No. | $t$ | $x$ | $\dot{x}$ | $Q(x)$. | $m(x)$ | $p_x$ |
|---|---|---|---|---|---|---|
| 1 | 0 | 1 | 0 | 0.14984 | 0.75026 | 0 |
| 2 | 9.41004E-5 | 1 | −8E-5 | 0.14984 | 0.75026 | −6E-5 |
| 3 | 5.64602E-4 | 1 | −4.8E-4 | 0.14984 | 0.75026 | −3.6E-4 |
| 4 | 0.00292 | 1 | −0.00248 | 0.14984 | 0.75026 | −0.00186 |
| 5 | 0.01468 | 0.99991 | −0.01248 | 0.14985 | 0.75024 | −0.00936 |
| 6 | 0.07349 | 0.9977 | −0.06242 | 0.1501 | 0.74974 | −0.0468 |
| 7 | 0.19277 | 0.98425 | −0.16281 | 0.15158 | 0.74671 | −0.12157 |
| 8 | 0.35773 | 0.94619 | −0.29727 | 0.15578 | 0.73802 | −0.21939 |
| 9 | 0.56268 | 0.86908 | −0.45197 | 0.16432 | 0.72003 | −0.32543 |

**Table A1** (*continued*)

| Sl. No. | t | x | ẋ | Q(x). | m(x) | $p_x$ |
|---|---|---|---|---|---|---|
| 10 | 0.8076 | 0.73825 | −0.61014 | 0.17879 | 0.68842 | −0.42003 |
| 11 | 1.10091 | 0.53734 | −0.7489 | 0.20051 | 0.63789 | −0.47772 |
| 12 | 1.38849 | 0.30989 | −0.82162 | 0.22344 | 0.5792 | −0.47588 |
| 13 | 1.64128 | 0.10004 | −0.83008 | 0.24213 | 0.52524 | −0.43599 |
| 14 | 1.78967 | −0.02192 | −0.81097 | 0.25163 | 0.49453 | −0.40105 |
| 15 | 1.93806 | −0.13977 | −0.77465 | 0.25973 | 0.4656 | −0.36068 |
| 16 | 2.10593 | −0.26498 | −0.714 | 0.26712 | 0.43587 | −0.31121 |
| 17 | 2.32048 | −0.40736 | −0.60902 | 0.27394 | 0.40353 | −0.24576 |
| 18 | 2.58348 | −0.54654 | −0.44499 | 0.27895 | 0.37363 | −0.16626 |
| 19 | 2.86371 | −0.64272 | −0.23898 | 0.28147 | 0.35402 | −0.08461 |
| 20 | 3.10574 | −0.67728 | −0.04656 | 0.28219 | 0.34719 | −0.01616 |
| 21 | 3.25751 | −0.67501 | 0.07604 | 0.28215 | 0.34764 | 0.02644 |
| 22 | 3.40929 | −0.65423 | 0.197 | 0.28172 | 0.35173 | 0.06929 |
| 23 | 3.59514 | −0.60423 | 0.33906 | 0.28055 | 0.36176 | 0.12266 |
| 24 | 3.82111 | −0.50932 | 0.4966 | 0.27777 | 0.38145 | 0.18943 |
| 25 | 4.08596 | −0.35637 | 0.65049 | 0.27169 | 0.41492 | 0.2699 |
| 26 | 4.35082 | −0.16817 | 0.76142 | 0.26152 | 0.45876 | 0.34931 |
| 27 | 4.52112 | −0.03435 | 0.80632 | 0.25253 | 0.49144 | 0.39626 |
| 28 | 4.69141 | 0.10522 | 0.82871 | 0.2417 | 0.52656 | 0.43636 |
| 29 | 4.83679 | 0.22596 | 0.82932 | 0.23124 | 0.55751 | 0.46235 |
| 30 | 5.0241 | 0.37944 | 0.80462 | 0.21668 | 0.59721 | 0.48052 |
| 31 | 5.24693 | 0.55211 | 0.7385 | 0.19895 | 0.64166 | 0.47387 |
| 32 | 5.50401 | 0.7272 | 0.61608 | 0.18001 | 0.6857 | 0.42244 |
| 33 | 5.7995 | 0.88195 | 0.42394 | 0.16289 | 0.72307 | 0.30654 |
| 34 | 6.07287 | 0.96924 | 0.21135 | 0.15323 | 0.7433 | 0.1571 |
| 35 | 6.2513 | 0.99372 | 0.06257 | 0.15054 | 0.74885 | 0.04685 |
| 36 | 6.3851 | 0.99451 | −0.05066 | 0.15045 | 0.74902 | −0.03794 |
| 37 | 6.51889 | 0.98019 | −0.16292 | 0.15202 | 0.74579 | −0.1215 |
| 38 | 6.68406 | 0.94209 | −0.297 | 0.15623 | 0.73708 | −0.21891 |
| 39 | 6.88928 | 0.86499 | −0.45123 | 0.16477 | 0.71906 | −0.32446 |
| 40 | 7.1345 | 0.73425 | −0.60881 | 0.17923 | 0.68744 | −0.41852 |
| 41 | 7.42818 | 0.53357 | −0.74691 | 0.20091 | 0.63692 | −0.47572 |
| 42 | 7.71554 | 0.30698 | −0.81893 | 0.22372 | 0.57845 | −0.47371 |
| 43 | 7.96804 | 0.0981 | −0.82707 | 0.24229 | 0.52475 | −0.434 |
| 44 | 8.11563 | −0.02277 | −0.80802 | 0.25169 | 0.49432 | −0.39942 |
| 45 | 8.26323 | −0.13957 | −0.77202 | 0.25972 | 0.46565 | −0.35949 |
| 46 | 8.43119 | −0.26442 | −0.71157 | 0.26709 | 0.436 | −0.31024 |
| 47 | 8.64579 | −0.40635 | −0.60695 | 0.27389 | 0.40376 | −0.24506 |
| 48 | 8.90881 | −0.54508 | −0.44347 | 0.27891 | 0.37394 | −0.16583 |
| 49 | 9.189 | −0.64091 | −0.23819 | 0.28143 | 0.35438 | −0.08441 |
| 50 | 9.43098 | −0.67535 | −0.04643 | 0.28215 | 0.34757 | −0.01614 |
| 51 | 9.58271 | −0.6731 | 0.07572 | 0.28211 | 0.34801 | 0.02635 |
| 52 | 9.73444 | −0.65241 | 0.19624 | 0.28168 | 0.35209 | 0.0691 |
| 53 | 9.92024 | −0.60261 | 0.3378 | 0.28051 | 0.36209 | 0.12231 |
| 54 | 10 | −0.57335 | 0.39547 | 0.27973 | 0.36808 | 0.14557 |

.